\documentclass[aps,pra,twocolumn,showpacs]{revtex4}
\usepackage{psfig}
\begin{document}

\title{Conditional generation of $N$-photon entangled states of light}

\author{Jarom\'{\i}r Fiur\'{a}\v{s}ek}

\affiliation{Department of Optics, Palack\'{y} University, 17. listopadu 50,
77200 Olomouc, Czech Republic}

\begin{abstract}
We propose a scheme for conditional generation of two-mode
$N$-photon path-entangled states of traveling light field.
These states may find applications in quantum optical lithography
and they may be used to improve the sensitivity of interferometric
measurements. Our method requires only single-photon sources,
linear optics (beam splitters and phase shifters), and photodetectors
with single photon sensitivity.

\end{abstract}


\maketitle

Quantum entanglement represents one of the most remarkable and
intriguing features of the quantum mechanics. Recently, entanglement has
been identified as a fundamental resource necessary for quantum
information processing \cite{Bouwmeester00} such as quantum teleportation
\cite{Bennett93,Zeilinger97,Braunstein98,Furusawa98}
and quantum computing \cite{Steane98}.
The entangled states may also help to improve
the sensitivity of interferometric measurements
\cite{Yurke86,Hillery93,Brif96,Dowling98}
and they form a key ingredient of quantum optical lithography
\cite{Boto00,Bjork00,DAngelo01},
which employs $N$-photon entangled states to fabricate patterns on
lithographic substrate with resolution $\lambda/(2N)$,
where $\lambda$ is the optical wavelength.

In view of these potential applications, it is highly desirable
to build a source of $N$-photon path-entangled states of traveling light
field,
\begin{equation}
\vert \psi_N\rangle =\sum_{k=0}^N c_k\vert k,N-k\rangle.
\label{TARGET}
\end{equation}
Here  $\vert k,N-k\rangle$ denotes the usual Fock state with $k$ photons in
mode $a$ and $N-k$ photons in mode $b$. Of particular interest could be
the entangled state
\begin{equation}
\vert \psi_N^0\rangle=\frac{1}{\sqrt{2}}\left(\vert N,0\rangle +
\vert 0,N\rangle \right).
\label{N}
\end{equation}
In Schr\"{o}dinger picture, $\vert \psi_N^0\rangle$ evolves in time according
to $\vert \psi_N^0(t)\rangle=\exp(-iN \omega t)\vert \psi_N^0(0)\rangle$,
where $\omega=2\pi c/\lambda$. We can interprete (\ref{N}) as a
state of a quasiparticle with energy $N\hbar \omega$
and effective de-Broglie wavelength $\lambda_{\rm eff}=\lambda/N$. This
specific feature of $\vert \psi_N^0\rangle$ is the origin of the
improvement of the resolution in quantum optical lithography
\cite{Boto00,Bjork00,DAngelo01}.

For $N=2$, the state (\ref{N}) can be generated by feeding two-ports
of a balanced beam splitter with  single-photon Fock states,
e.g., signal and idler photons generated by means of spontaneous
parametric down-conversion \cite{Mandel87}. For $N>2$, however,
single-photon sources and linear optics are not sufficient for
deterministic preparation of state (\ref{N}), and Kerr or other
nonlinear media are required. Unfortunately,
sufficiently strong nonlinear interactions between single traveling
photons are not currently available.

Nevertheless, one may avoid the necessity of nonlinear interactions.
In a recent paper, Lee {\em et al.} \cite{Lee01} showed that
the states (\ref{N}) can be prepared probabilistically using
only Fock-state sources,
linear optical elements and single-photon counting detectors. Lee {\em
et al.} provided schemes for $N=3$ and $N=4$ but were not able
to extend them to higher $N$. In the present paper we design
scheme for conditional generation of {\em arbitrary} entangled
 $N$-photon states (\ref{TARGET})
for {\em any} $N$. We first present a generic scheme and then,
as an application, we shall consider preparation of the state (\ref{N}).

The  quantum-state preparation schemes whose success is conditioned on
the results of quantum measurements have attracted
considerable amount of attention recently.
Schemes for probabilistic preparation of
Fock states \cite{Cirac93,DAriano00},
arbitrary superpositions of Fock states of single-mode
field \cite{Vogel93,Dakna99}, and
Schr\"{o}dinger cat states \cite{Song90,Dakna97},
have been found. Experimental conditional preparation of the single-photon
Fock state with negative Wigner function has been reported \cite{Lvovsky01}.
In cavity QED, the state (\ref{TARGET}) can be generated by
injection of a sequence of $N$ suitably prepared three-level $\Lambda$-type
atoms into a two-mode resonator \cite{Deb95}. In that scheme, one
detects whether the atom leaving the resonator is in excited or ground
state and the desired state (\ref{TARGET}) is prepared only  if
all atoms are in a ground state.
Here, we design scheme for generation of
two-mode entangled states of {\em traveling} light field.

For our purposes it is convenient to express the target state
(\ref{TARGET}) in terms of bosonic creation operators $a^\dagger$ and
$b^\dagger$ acting on two-mode vacuum state,
\begin{equation}
\vert \psi_N\rangle=\sum_{k=0}^N d_k a^{\dagger k} b^{\dagger N-k}
\vert 0,0\rangle,
\label{TARGETAB}
\end{equation}
where $d_k=c_k/\sqrt{k!(N-k)!}$.
The  polynomial on the right-hand side of Eq. (\ref{TARGETAB}) can be
factorized into a product of $N$ terms linear in creation operators,
\begin{equation}
\vert \psi_N\rangle =\frac{1}{\sqrt{\cal{N}}}
\prod_{k=1}^N \left(\cos\theta_ka^{\dagger}
-e^{i\phi_k}\sin\theta_k b^{\dagger}\right) \vert 0,0\rangle,
\label{FACTORED}
\end{equation}
where $\cal{N}$ is a normalization factor and
$z_k=e^{i\phi_k}\tan\theta_k$  are complex roots of the polynomial
$\sum_{k=0}^n d_k z^k.$
The factorization (\ref{FACTORED}) suggests that we can prepare
the state $\vert \psi_N\rangle$ from the vacuum state $\vert 0,0\rangle$
by applying $N$-times a non-unitary transformation
\begin{equation}
\vert \psi_{k}\rangle = \left(
\cos\theta_ka^{\dagger}
-e^{i\phi_k} \sin\theta_k b^{\dagger}\right) \vert \psi_{k-1}\rangle,
\label{BLOCK}
\end{equation}
starting with $\vert \psi_0\rangle=\vert 0,0\rangle$.
We show that the transformation (\ref{BLOCK}) can be implemented
probabilistically using only single-photon sources, linear optical elements
and photodetectors, provided that $\vert \psi_k\rangle$ is an eigenstate
of the operator of total number of photons
$n_{ab}=a^\dagger a+b^{\dagger}b$,
\begin{equation}
n_{ab}\vert \psi_k\rangle = k \vert \psi_k\rangle.
\label{EIGEN}
\end{equation}

The setup under consideration is shown in Fig. 1. In addition to modes
$a$ and $b$, which contain the state $\vert \psi\rangle$, we also need
two auxiliary modes $c$ and $d$, initially in a single-photon Fock
state $\vert 1,0\rangle_{cd}$. With the help of the beam splitter $BS_1$
with transmittance $\cos\theta_k$ and a suitable phase shifter
we transform this state to
\begin{equation}
\vert \varphi_k\rangle_{cd}=\cos\theta_k\vert 1,0\rangle_{cd}
-e^{i\phi_k}\sin\theta_k\vert 0,1\rangle_{cd}.
\label{FICD}
\end{equation}
The next step consists of mixing the mode $a$ with $c$ at a beam
splitter BS$_2$ while the  mode $b$ is mixed with $d$ at BS$_3$.
The beam splitters BS$_2$ and BS$_3$ are identical.
The corresponding unitary transformation describing the operation of
BS$_2$ and BS$_3$ can be thus parametrized by a single real number
$\kappa$,
\begin{equation}
U=\exp(\kappa a^\dagger c- \kappa a c^\dagger)\,
\exp(\kappa b^\dagger d- \kappa b d^\dagger).
\label{U}
\end{equation}
The photodetectors PD$_1$ and PD$_2$ measure number of photons in the
output modes $c_{\rm out}$ and $d_{\rm out}$.
In our scheme the operation (\ref{BLOCK}) is
successfully applied if and only if both detectors do not detect any
photons. This means that PD$_j$ need not resolve between single and
two-photon states, they should only  distinguish vacuum state from any
Fock state with nonzero number of photons. Efficient avalanche
photodiods are suitable for this purpose \cite{Kwiat93}.

\begin{figure}[!t!]
\centerline{\psfig{figure=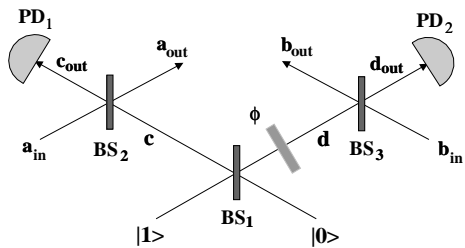,width=0.7\linewidth}}
\caption{Setup for probabilistic implementation of the transformation
(5) consists of single photon source, three beam-splitters  BS$_1$,
 BS$_2$, BS$_3$, phase shifter $\phi$, and two photodetectors
 PD$_1$ and PD$_2$.}
\end{figure}

If no photons are detected in output modes $c_{\rm out}$ and $d_{\rm out}$,
then the photon contained in the input state $\vert \varphi_k\rangle_{cd}$
has been added to modes $a$ or $b$. This intuitively explains the principle of
operation of the scheme shown in Fig. 1. In order to provide a rigorous
mathematical treatment, we rewrite the unitary transformation
(\ref{U}) in a disentangled form \cite{Collet88},
\begin{equation}
U=e^{-K a c^\dagger}e^{-K b d^\dagger}
(\cos \kappa)^{n_{ab}-n_{cd}}
e^{K a^\dagger c}
e^{K b^\dagger d},
\label{UFACTORED}
\end{equation}
where $K=\tan\kappa$ and $n_{cd}=c^\dagger c+d^\dagger d$.
The conditionally generated  state $\vert \psi_{k}\rangle$
in the output modes $a_{\rm out}$ and $b_{\rm out}$ can be obtained
from the transformed input state
$U \vert \psi_{k-1}\rangle_{ab}\vert \varphi_k\rangle_{cd}$
by applying a projection operator
\begin{equation}
{\Pi}= \openone_{ab} \otimes \vert 0,0\rangle_{cd}\langle 0,0\vert ,
\label{PI}
\end{equation}
which describes the conditioning on no photons present in the
modes $c_{\rm out}$ and $d_{\rm out}$.
Here $\openone_{ab}$ is an identity operator
acting on Hilbert space of modes $a$ and $b$.
Thus we can write
\begin{equation}
\vert \psi_{k}\rangle_{ab}\vert 0,0\rangle_{cd}=
{\Pi} \,{U} \,\vert \psi_{k-1}\rangle_{ab}\vert \varphi_k\rangle_{cd}.
\label{PSIK}
\end{equation}
We insert the factorized form (\ref{UFACTORED}) of the operator $U$
into Eq. (\ref{PSIK}) and make use of the vacuum stability condition
$_{cd}\langle 0,0\vert  c^\dagger =\!\!~_{cd}\langle 0,0\vert  d^\dagger=0$
to simplify Eq. (\ref{PSIK}) as follows,
\begin{equation}
\vert \psi_{k}\rangle_{ab}\vert 0,0\rangle_{cd}=
\Pi \, (\cos \kappa)^{n_{ab}} e^{K a^\dagger c} e^{K b^\dagger d}
\vert \psi_{k-1}\rangle_{ab}\vert \varphi_k\rangle_{cd}.
\label{PSIKSIMPLE}
\end{equation}
Now we expand the two exponentials in Taylor series. Since the
state $\vert \varphi_k\rangle_{cd}$ contains only a single photon, we have to
keep only terms up to linear in annihilation operators $c$ and $d$,
\begin{equation}
\exp(K a^\dagger c+ Kb^\dagger d) \rightarrow 1+K a^\dagger c +K b^\dagger d.
\label{TAYLOR}
\end{equation}
Inserting (\ref{TAYLOR}) back into Eq. (\ref{PSIKSIMPLE}) and taking
into account that $_{cd}\langle 0,0\vert \varphi_k\rangle_{cd}=0$ we obtain
\begin{equation}
\vert \psi_{k}\rangle_{ab}= q_k
(\cos\theta_k a^\dagger - e^{i\phi_k}\sin\theta_k b^\dagger)
\vert \psi_{k-1}\rangle_{ab},
\label{PSIKFINAL}
\end{equation}
where
\begin{equation}
q_k= (\cos\kappa)^{k-1} \sin\kappa
\label{QK}
\end{equation}
and we have used that the state $\vert \psi_{k-1}\rangle$ is an
eigenstate of the total photon number operator $n_{ab}$.

The desired $N$-photon entangled state (\ref{TARGET}) can be generated if we
repeatedly $N$-times apply the basic transformation (\ref{BLOCK}),
as illustrated in Fig. 2. There are altogether $2N$ detectors.
If none of them detects any
photon, then the state $\vert \psi_N\rangle$ is generated at the output.
We assume that $\kappa$ may be different for each basic
building block, hence we have $3N$ parameters $\kappa_k$, $\theta_k$ and
$\phi_k$ ($k=1,\ldots,N$) characterizing the setup shown in Fig. 2.
The unnormalized conditionally generated output state reads
\begin{equation}
\vert \psi_N\rangle= \prod_{k=1}^{N} q_{k} (\cos\theta_k a^\dagger
 -e^{i\phi_k}\sin\theta_k b^\dagger) \vert 0,0\rangle.
\label{PSIFINAL}
\end{equation}
The probability $P_N$ of generation of the state $\vert \psi_N\rangle$,
i.e., the yield of our scheme, can be obtained as a norm of the
output state (\ref{PSIFINAL}),
\begin{equation}
P_N= {\cal{N}} \prod_{k=1}^N q_k^2.
\label{PN}
\end{equation}
We can maximize the probability $P_N$ by maximizing independently
each term $q_k^2$. It is convenient to introduce a transmittance
$T_k=\sin^2\kappa_k$. Thus we have
\begin{equation}
q_k^2= T_k(1-T_k)^{k-1}
\label{QKT}
\end{equation}
and the optimal $T_k$ maximizing $q_k^2$ reads
\begin{equation}
T_k=\frac{1}{k}.
\label{TKOPT}
\end{equation}
Notice that the optimum beam-splitter transmittance does not depend on
the state which we want to generate.
On inserting $q_k^2=(k-1)^{k-1}/k^{k}$  into Eq. (\ref{PN}) we obtain
the optimum probability of generation
\begin{equation}
P_N= {\cal{N}} N^{-N}.
\label{PNOPT}
\end{equation}
The normalization factor $\cal{N}$ has been introduced in Eq.
(\ref{FACTORED}).

As an example of application of our generic method, we shall consider
generation  of the entangled state (\ref{N}). It is easy to see that
this state may be written as follows,
\begin{equation}
\vert \psi_N^0\rangle=\frac{1}{\sqrt{2}\sqrt{N!}}\prod_{k=1}^N
(a^\dagger-e^{i\phi_k} b^\dagger)\vert 0,0\rangle,
\label{PSINFACTORED}
\end{equation}
where $\phi_k=(2k+1)\pi/N.$
Upon comparing Eqs. (\ref{PSINFACTORED}) and (\ref{FACTORED})
we find that  $\theta_k=\pi/4$. After some algebra one obtains
the optimum probability of generation
\begin{equation}
P_{N}=(N-1)! \, (2N)^{1-N}.
\label{POPTN}
\end{equation}
With the help of Stirling's formula we find that for large $N$ we may
approximate Eq. (\ref{POPTN}) as $P_N \approx 2\sqrt{2\pi N} (2e)^{-N}$.
The yield decays exponentially with the number of photons $N$.

\begin{figure}[t]
\centerline{\psfig{figure=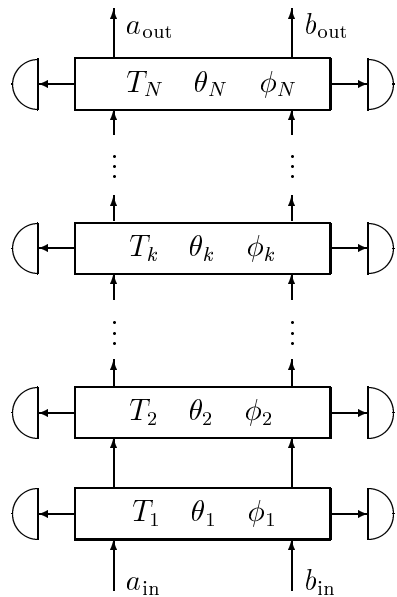,width=0.6\linewidth}}

\caption{Setup for probabilistic generation of $N$-photon path-entangled
states of light. The device consists of $N$ basic building blocks,
each block represents the scheme shown in Fig. 1 and is characterized
by three parameters $T_k$, $\theta_k$, and $\phi_k$.
}
\end{figure}

If $N$ is even, then we can simplify our scheme and reduce the
number of necessary elements by a factor of two.
We write the state (\ref{N}) as follows,
\begin{equation}
\vert \psi_N^0\rangle= \frac{1}{\sqrt{2}\sqrt{N!}}\prod_{k=1}^{N/2}
(a^{\dagger 2}-e^{2i\phi_k}b^{\dagger 2}) \vert 0,0\rangle.
\end{equation}
We can generate this state if we perform $N/2$-times the transformation
\begin{equation}
\vert \psi_k\rangle = (a^{\dagger 2}-e^{2i\phi_k}b^{\dagger 2})
\vert \psi_{k-1}\rangle,
\end{equation}
which can be conditionally implemented with only a slight modification
of the scheme shown in Fig. 1. Instead of the vacuum state $\vert 0\rangle$,
we send a single-photon Fock state $\vert 1\rangle$ into the right
input port of $BS_1$. After mixing on balanced
beam-splitter $BS_1$ [$\theta=\pi/4$] and passing through the phase-shifter,
the state $\vert \varphi_k\rangle_{cd}$ of the modes $c$ and $d$ reads
\begin{equation}
\vert \varphi_k\rangle_{cd} =\frac{1}{\sqrt{2}}\left(\vert 2,0\rangle_{cd}
                      -e^{2i\phi_k}\vert 0,2\rangle_{cd}\right).
\end{equation}
Similarly as before, we condition on detecting no photons
in the output modes $c_{\rm out}$ and $d_{\rm out}$. In this way we add two
photons to the modes $a$ or $b$ at each step.
After $N/2$ steps we thus end up with $N$-photon entangled state.

The calculations of the conditionally generated output state closely
follow those presented above. Since the state
$\vert \varphi_k\rangle_{cd}$ now contains two photons,
we must keep quadratic terms in the expansion (\ref{TAYLOR}),
\begin{eqnarray}
e^{K a^\dagger c} e^{K b^\dagger d} &\rightarrow&
1+K(a^\dagger c +b^\dagger d)
\nonumber \\
&&+\frac{K^2}{2}(a^{\dagger 2} c^{2}+2a^\dagger b^\dagger c d
+b^{\dagger 2}d^2).
\label{TAYLORTWO}
\end{eqnarray}
Assuming that the state $\vert \psi_{k}\rangle$
is an eigenstate of total number of photons,  $n_{ab}\vert \psi_{\rm
k}\rangle=2k\vert \psi_k\rangle$, we find that
\begin{equation}
\vert \psi_{k}\rangle =\frac{1}{2}(\cos\kappa)^{2k}\tan^2\kappa
(a^{\dagger 2} - e^{2i\phi_k}b^{\dagger 2})\vert \psi_{k-1}\rangle.
\end{equation}
The optimal transmittance of the $k$-th beam-splitter is again
given by Eq. (\ref{TKOPT}). The probability of generation of
the state (\ref{N}) (i.e., the yield) reads
\begin{equation}
P_N^\prime= 2 \, (N-1)!\, N^{1-N}
\label{POPTNEVEN}
\end{equation}
A comparison of the yields (\ref{POPTNEVEN}) and (\ref{POPTN})
immediately reveals that $P_N^\prime= 2^N P_N$.
The scheme where we add two photons in a single step is much more
effective, because the number of necessary measurements is halved.
To be specific, for $N=4$ we have $P_4=3/256$ and $P_4^\prime=3/16$.
Lee {\em et al.} \cite{Lee01} designed schemes for
generation of the state $(\vert 4,0\rangle+\vert 0,4\rangle)/\sqrt{2}$ with yield
$3/64$ and our second method improves on this result by a factor of 4.

On the way towards experimental implementation of the scheme proposed in
the present paper, two main obstacles have to be overcome.
First, one needs a controlled source of single-photon Fock states.
Currently available triggered single photon
sources operate by means of fluorescence from a single molecule \cite{Brunel99}
or a single quantum dot \cite{Michler00,Santori01} and they exhibit
very good performance. However, in our scheme, we need a synchronized
arrival of $N$ single photons into $N$ input ports of $N$
beam splitters, which will be experimentally challenging.

A second obstacle stems from the
less-than-unit efficiencies of the single-photon detectors.
Imperfect detectors  will degrade the output state which will be a
mixed state described by some density matrix $\rho_{ab}$ \cite{Lee01}.
However, in some applications, such as quantum
lithography, this problem may be circumvented
because the detectors are actually not necessary.  If no conditioning
is performed, then the mixed output state can be expressed as
\vspace*{-1mm}
\begin{equation}
\rho_{ab}= P_{N}\vert \psi_N\rangle\langle \psi_{N}\vert
+(1-P_N)\tilde{\rho}_{ab},
\end{equation}
where the density matrix $\tilde{\rho}_{ab}$ represents the output
state when one or more photons leak into the output auxiliary modes.
This implies that the operator $\tilde{\rho}_{ab}$ is supported on Hilbert
space of Fock states $\vert k,M-k\rangle$ with $M\leq N-1$.

Now consider the quantum lithography.
If the lithographic process is based on the $N$-photon absorption, then
the absorption rate at the imaging surface will be proportional to the
expectation value of  normally ordered operator \cite{Boto00}
\vspace*{-4mm}
\begin{equation}
\delta=\frac{e^{\dagger N} e^N}{N!},
\end{equation}
where $e=a+b$ is the effective positive-frequency field operator.
It follows that the medium will respond only to the $N$-photon part
of the output state $\rho_{ab}$ and we have
\begin{equation}
{\rm Tr} \rho_{ab} \delta= P_N \langle\psi_N\vert \delta\vert \psi_N\rangle,
\end{equation}
which is essentially the same result as for an ideal pure output state
$\vert \psi_N\rangle\langle\psi_N\vert $. The rate is only reduced by  factor
$P_N$ representing the yield of our scheme.

In summary, we have designed a universal scheme for conditional
generation of an arbitrary $N$-photon path-entangled quantum state
of traveling light field. The necessary resources comprise
single photon sources, beam splitters, phase shifters, and
photodetectors with single-photon sensitivity. However,
in certain applications, when one wishes to measure or utilize
the $N$-photon coincidence rates, the conditioning is not necessary,
because the desired $N$-photon part of the output state is selected
automatically.

The author would like to thank N.J. Cerf, S. Iblisdir, and R. Filip for
useful discussions. This work was supported by Grant No LN00A015 and
Research Project CEZ: J14/98 of the Czech Ministry of Education.

\end{document}